# Teaching Computation in Introductory Physics using Complex Problems

Marcos D. Caballero[1,2,3], Michael J. Obsniuk[1,4], Paul W. Irving[1]
[1] Department of Physics and Astronomy, Michigan State University
[2] CREATE for STEM Institute, Michigan State University
[3] Department of Physics and Center for Computing in Science Education, University of Oslo
[4] Department of Physics, Kettering University

## Abstract

Computation is a central aspect of modern science and engineering work, and yet, computational instruction has yet to fully pervade university STEM curricula. In physics, we have begun to integrate computation into our courses in a variety of ways. Here, we discuss a method for integrating computation into calculus-based mechanics where the lecture and laboratory for the course are decoupled. At Michigan State University, we have developed a "lecture" course, called "Projects and Practices in Physics", where science and engineering students solve complex problems in groups of four using analytical and computational techniques. In this paper, we provide details on the computational instruction, activities, and assessment used to teach these introductory students how to model motion using VPython.

## Introduction and Background

Numerous calls have been made to modernize science curricula to emphasize that students should learn the practices of professional science [1-5]. These calls push us to better represent what science (and engineering) is and does in our courses. The aim is that students will develop deeper and more robust knowledge of science and that the participation in science will grow as students are actively engaged in the construction of their knowledge. One practice that is central to 21st century science is computational modeling – using a computer to solve, simulate, and/or visualize a problem. Computational modeling is so ubiquitous in professional science (and STEM, more broadly), it is considered central to the modern scientific endeavor along with theory and experiment. And yet, there is a relative absence of computational modeling in much of a student's undergraduate science experience.

While computation might be lacking in science course work broadly, physics instructors appear to be leading the way in computational instruction. A growing number of high school teachers and university educators have begun to incorporate computational modeling into their courses in a wide variety of ways. One of the most well-known and widely-visible examples is the Matter and Interactions (M&I) curriculum [6]. This curriculum is used by educators in high school, two-year college, and university settings across the country. At the university level, it is typical to teach computational modeling using M&I in the laboratory – leveraging the small student-to-instructor ratio and group-focused instruction that is typical of laboratory instruction. Introducing computation in this way can stem from a number of different goals including immersing students in making models of physical systems [7], developing students problem-solving skills [8], and helping students to make sense of different phenomenon [9]. There's a wide breadth to how computational physics has been used in these courses. In some cases, the computational experiences students gain in the laboratory have been leveraged on homework [8]. In other cases, students have used computation as a means to understanding motion in their everyday life [10,11].Still in others, non-science majors have used computation to recreate video games – accurately modeling the essential physics to make those games work (cite Titus). In still others, students have worked on semester long computational physics projects (cite Dave and Dwain's work). The use of computational physics in a variety of spaces is encouraging as instructors work to determine which model of computational instruction best fits with their particular needs and situational factors.

In this paper, we provide another method of computational instruction that is conducted in a group-based introductory mechanics course with no laboratory [12]. The work we describe makes use of a flipped model of instruction [13,14], context-rich problems inspired by problem-based learning models [15,16], and minimal working programs [9]. We provide details on the context and learning goals for our course, offer a description of how computation is taught in the course including a detailed example, and describe what we think this model appears to be buying us and why you might consider adopting/adapting it.

## Context and Learning Goals

At Michigan State University, lectures courses are divorced from laboratory courses – students might take the laboratory at the same time, at a later time, or not at all. This makes incorporating computation into laboratory courses problematic as most models for teaching computation in the laboratory make use of a courses that are deeply coupled to the lecture course [17] or in studio-style settings [18]. Our typical lecture courses are taught in a large-lecture hall with fixed seating and have 275 students lead by a single instructor. This model of instruction is not amenable to teaching computation as it provides little opportunity for students to engage with computation in the scaffolded and supportive manner needed to support all students [8,9]. We wanted to explore a different model of instruction [12] that reduces class size to 100 students – similar to SCALE-UP and Modeling Instruction classrooms, emphasizes practices – including computational modeling, and does not rely on laboratory course coupling.

The course we developed called "Projects and Practices in Physics" (or $P^3$) is an introductory mechanics course where students work in groups of four to develop solutions to open-ended, messy, context-rich problems that we refer to as "projects" [12]. Student work is supported by pre-class readings and lecture videos as well as pre-class homework that is meant to "prime the pump" for class meetings. In class, students work in their groups to develop their solutions by identifying what they know from the problem statement, what they need to learn more about, what principles and models apply to the situation, and what approach they should take. While $P^3$ is not entirely a problem-based learning course, it is heavily inspired by problem-based learning. $P^3$ emphasizes the three conservation principles of mechanics (linear momentum, energy, and angular momentum) and it focuses students on a first principles approach. $P^3$ is structured around the Matter and Interactions textbook [6], but follows an approach that fits with the situational factors at Michigan State (i.e., no coupled laboratory). More details about the course are available in a recent publication [12]. This paper focuses on the specifics of computational instruction in this mechanics course.

Because $P^3$ is a general university-level mechanics course for scientists and engineers, our computational learning goals emphasize an introduction to predicting motion iteratively using the Euler-Cromer step. We have three principal learning goals for computation modeling in this course:
- LG 1 - Students should be able to identify and interpret initial conditions and model parameters, make changes to them as needed, and include additional parameters as necessary. This goal emphasizes that students should be able to make physical meaning of the program they are using and editing. Here, students must make sense of variables assignment and types (e.g., number, vector, string) as well as operations that can be conducted on those variables.
- LG 2 - Students should be able to construct the integration loop including the force model and the numerical integration (Euler-Cromer) step for a wide variety of phenomenon. This goal deals with the intersection of control structures (the loop and its control variable) and the underlying physics (step-wise integration of the equations of motion). There's some additional nuance here in that some of the later problems (Ex 2) include new control structures (like if statements) and rotational models (torque and angular momentum).
- LG 3 - Students should begin to see the utility of computational modeling as a modern approach to doing physics. This goal is not measurable in the traditional sense, although, we have several research projects that begin to unpack how students perceive the use of computation in this course and science and engineering more broadly.

It should be noted that none of these goals indicate that students should (necessarily) construct the computational modeling from scratch. This was a decision made given our student population - less than 10% of students taking $P^3$ have had any prior computational modeling experience - and the nature of the course - $P^3$ is an introductory physics with computation not an introductory computational physics course.

## How is computational modeling taught?

Computational instruction emphasizes the model for instruction used in other aspects of the course [12]. Students read and solve pre-class homework questions that are meant to prepare them to work on open-ended (computational) problems in groups of four during class. Students are given formative feedback from their tutors on their approach to solving these problems and how well they are working in their group. After completing their classwork, they solve some more traditional post-class homework drawn from the M&I textbook [19].

Consider the following VPython script that was written to model the motion of a fan cart.
*You may assume the script uses SI units.*

```
from visual import *

fancart = sphere(pos=vector(15.0,14.2,0), radius=0.5, color=color.red)
mcart = 3.61
vcart = vector(6.57,0,0)
pcart = mcart*vcart

Ffan = vector(3.71,0,0)

t = 0
dt = 0.1
tf = 3.90

while t < tf:
    rate(150)

    pcart = pcart + Ffan*dt
    fancart.pos = fancart.pos + (pcart/mcart)*dt

    t = t + dt
```

(a) At what location is the fan cart when the script starts?

$\vec{r}_i = \langle \boxed{\phantom{X}}, \boxed{\phantom{X}}, \boxed{\phantom{X}} \rangle$ m

(b) What is the velocity of the fan cart when the script starts?

$\vec{v}_i = \langle \boxed{\phantom{X}}, \boxed{\phantom{X}}, \boxed{\phantom{X}} \rangle$ m/s

(c) At what time does the script stop running?

$t_f = \boxed{\phantom{X}}$ s

(d) What is the momentum of the fan cart when the script stops running?

$\vec{p}_f = \langle \boxed{\phantom{X}}, \boxed{\phantom{X}}, \boxed{\phantom{X}} \rangle$ kg · m/s

*Figure 1 - Pre-class homework problem delivered prior to students' development of a computational model of constant force.*

Pre-class computational homework questions focus on preparing students to identify and to interpret already written code (LG1). In the example shown above (Figure 1), students are given a randomized piece of code and work to determine where the object is located and what velocity it has. Some questions given in future pre-class homework questions ask students to identify what kind of motion is being modeled (constant velocity, constant force, 1D vs. 2D, etc.) and how they can tell from the code that such motion is being modeled. By delivering these homework problems, we aim for students to gain some familiarity with the syntax of VPython and to recognize certain attributes (i.e., vectors) and structures (i.e., the while loop).

# Project 2: Part A: Escape from Ice Station McMurdo

You are a member of a scientific research team at McMurdo ice station which is funded by the Carver Media Group in Antarctica.

Two members of your research team have recently returned from investigating an incident at a Norwegian research facility. They brought with them a burnt humanoid body with two faces. Since the disturbing discovery several inhabitants of the ice station have disappeared. Frightened, a member of your team decided to flee the station on a fan powered hovercraft but you receive a distress call not long after their escape that their steering and acceleration controls have been jammed and they need your help.

| Time | Your Team Member's Position | Your Position |
|------|------------------------------|---------------|
| 0s   | 2536.40m                     | 10.47m        |
| 10s  | 3072.80m                     | 41.88m        |
| 20s  | 3609.20m                     | 94.22m        |

You decide to attempt a rescue in another hovercraft. You must decide how many members of your team help in the rescue operation. The hovercrafts do not have a velocity or acceleration gauge but they do have GPS locators and you possess your trusty stop watch. The GPS locator tells you the exact position of both your craft and other team members craft relative to the ice station. You are following their path. You collect the following data for the first 20 seconds of your journey.

You need to tell the runaway researcher the exact time from your starting time to jump onto your hovercraft as you may only have one shot at this rescue.

# Project 2: Part B: Escape from Ice Station McMurdo

Just as you are about to radio the time to jump to the runaway researcher, you realize the steering and acceleration controls have become frozen on your hovercraft and so it continues to accelerate and you cannot change direction. 200m ahead of the point at which you were going to tell the researcher to jump is an ice ravine. At the bottom of the ice ravine, 400m below, is an unfrozen salt water pool surrounded by stalagmites. From the ravine's edge to the pool is 490m and the pool stretches for 900m. You are moving too quickly to survive jumping off the hovercraft, but might survive the fall into the pool by staying on the hovercraft; it has seat belts. You now have a choice to make, to stay on your hovercraft or jump to the runaway researcher's hovercraft. One or both may make it to the pool. Your choice may be the difference between life and death..

# Project 2: Part C: Escape from Ice Station McMurdo

Surprisingly enough hovercrafts are an expensive piece of kit. Your employer, the Carver Media Group, is concerned by the happenings at the McMurdo ice station and would like you to produce an accident report detailing the events after you lost control of your hovercraft. The accident report should include a detailed computational model that provides the projected motion of the runaway hovercraft.

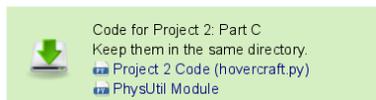

Code for Project 2: Part C
Keep them in the same directory.
📄 Project 2 Code (hovercraft.py)
📄 PhysUtil Module

*Figure 2 - Projects completed in class by students in groups of 4. Parts A and B are solved analytically while part C involves developing a computational solution.*

In class, students will be presented with complicated problems for which they must design a solution (LG2 and LG3). In the example above (Figure 2), students analyze a scenario where they must rescue a colleague whose hovercraft has gotten stuck in drive. In the first class period of the second week, students typically solve the first two parts of the problem, which are purely analytical problems - modeling constant force in 1 and 2 dimensions. In the very next class period, students are presented with an alternative result, that is, the hovercraft landed short - missing the safe landing zone - and they must file an incident report that attributes this reduced flight distance to something physical. The students easily suggest air resistance and set about doing a bit of research on how that effect is modeled, eventually finding Fair = -bv^2. Students are given a minimal working program that models the motion of the hovercraft up to the edge of the cliff.

By reading through the code and being guided to identify the different aspects of the Euler-Cromer integration step, students begin to write a second loop, which models the motion after the hovercraft leaves the cliff. Typically, students develop the model of the motion without air-resistance relatively quickly. From an expert perspective, doing so amounts to no more than copying the while loop, removing the force due to the surface, and changing the loop control variables to run until the hovercraft hits the ground. From the student perspective, each of these tasks involves some debate and discussion as they unpack what performing each of these tasks means, but they do recognize that first loop is quite similar to what they intend to do. The tutors do ask them to first focus on reproducing their work from the analytical day to check their model and to delay adding air resistance in until they have gotten a working bit of code. In a sense, we are teaching students to work with the simplest model possible (e.g., constant force) before adding complexity (e.g., air resistance)

By the end of the second class period, most of the groups are able to include air resistance into their code. From the student perspective, doing so is non-trivial, as the understanding of vectors needed to include air resistance is a large part of what students will learn over the entire course. Moreover, recognizing how those vectors are referenced in VPython and how they relate to the analytical formula is an exercise that requires coordination of different kinds of resources about physics, computation, and mathematics. Research has shown that students often encounter the difficulty with vectors in introductory physics [20-22] and that these concepts are similarly difficult in computational tasks [23].

Groups are encouraged to include conceptual aspects to their visualization, which are intended to develop their understanding of 2D motion and vectors. Students will include arrows that represent the x and y velocities. Tutors will ask questions about how these arrows change when air resistance is turned off – asking students to relate that to how forces change the motion. In addition, some groups might be asked to add graphs of these quantities to further unpack these conceptual learning goals. Below is an image of the resulting visualization that students produce.

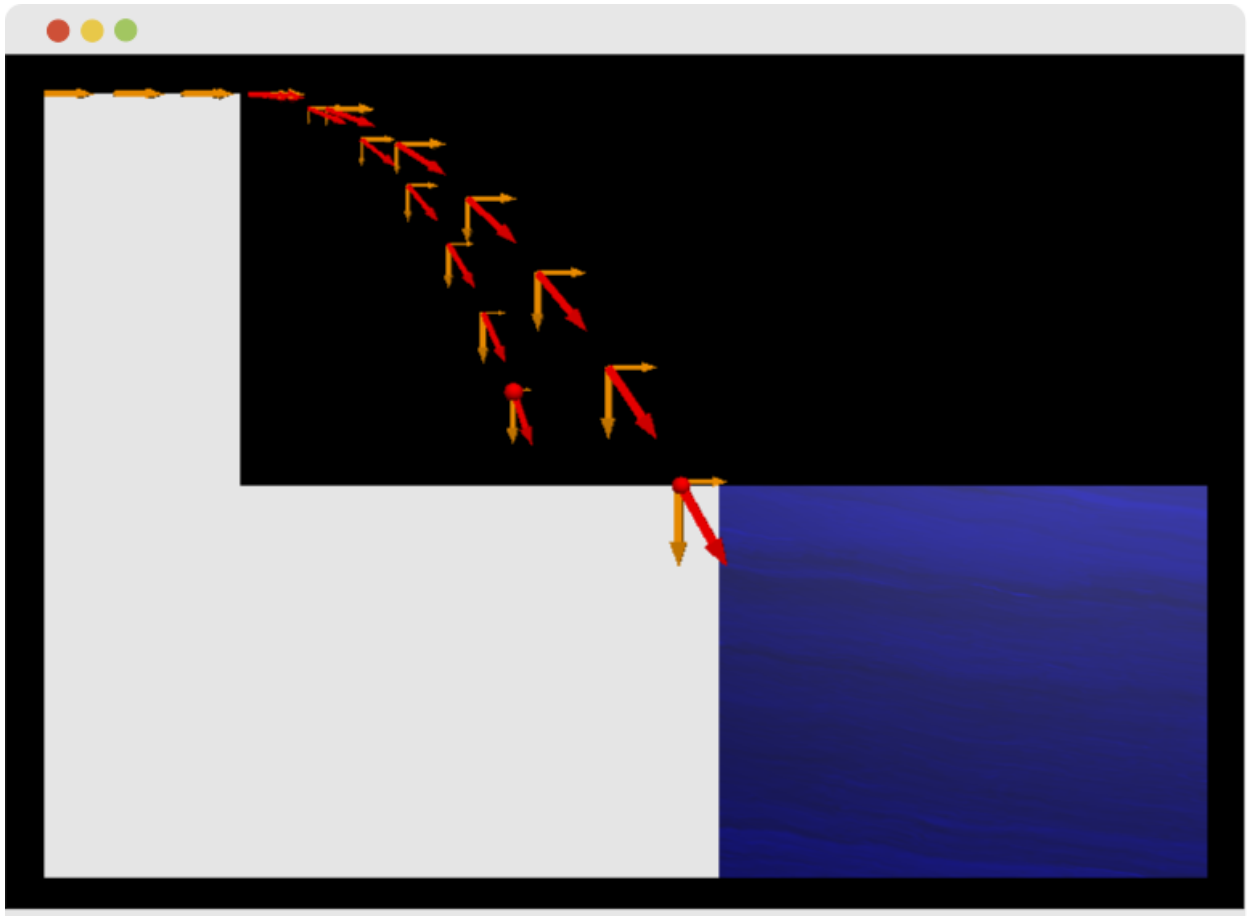

*Figure 3 - Student constructed VPython visualization of the project appears in Figure 2 including arrows representing the velocity vectors (red) and their components (orange).*

Over the course of the semester, we include computational modeling as part of 7 projects (out of 14 total). Below, we discuss briefly each project.
- Project 1: Modeling Relative and Constant Velocity: In this project, students model the motion of an asteroid and a satellite using a constant velocity model for each. The objective is for students to produce a graph of the relative separation of the two objects, which travel in 3 dimensions, to determine how far apart the objects are as a function of time. Students learn to write loops to predict the motion and produce graphs of quantities of interest.
- Project 2: Modeling Constant Force Motion in 1 and 2D (w/ Air Resistance): In this project (Figure 2), students model the motion of a hovercraft experiencing a constant force as it travels across a surface (in 1D) and falls off a cliff (in 2D). Students include a model for air resistance to observe how it affects the motion. Students produce arrows and graphs representing the horizontal and vertical velocity components to explore how they change in relation to the forces exerted.
- Project 3: Modeling Gravitationally Interacting Bodies (Geosynchronus Orbit): In this project, students model the motion of a satellite in orbit around the Earth. Students include arrows to represent the force and linear momentum. In addition, they investigate

- the quality of the simulation and its long-term behavior by plotting the relative separation distance between the Earth and the satellite as a function of time.
- Project 8: Investigating Energy (Launching a probe): In this project, students investigate how the energy of a gravitational system change with time and relative distance between the objects bound in the system. Students graph these energies and relate these graphs to arrows representing the parallel and perpendicular components of the net force on the probe.
- Project 11: Modeling Collisions (Protecting a space station): In this project, students develop a model of a collision between two objects in which the momentum is conserved by launching a projectile to intercept another projectile. Students develop an "if" statement to detect the collision and apply conservation of linear momentum to determine the velocities of both objects after the collision.
- Project 13: Modeling Rotation (Spinning disk): In this project, students develop a model of a collision between a projectile and extended system, which can rotate. Student develop an "if" statement to detect the collision and apply conservation of linear and angular momentum to determine the velocities of both objects and the rotation rate of the extended system.
- Project 14: Modeling with all 3 principles (Defense system): In this project, students develop a defense system, which relies on collisional dynamics and conservation principles to predict the motion of the defense projectile, which is an extended body.

Projects and other supporting materials are available online at pcubed.pa.msu.edu.

## Discussion and Conclusions

Preparing students to engage in modern scientific endeavors requires a computational approach to science education. Physics is at the forefront of this computational revolution. We are teaching computation in a wide variety of courses, formats, and levels. The approach discussed here is a way that emphasizes the inquiry-based approach that pervades other forms of physics instruction (ie., laboratories, tutorials, etc.) and has been shown to be highly productive for student learning in physics [24-26]. Scaffolds such as pre-class homework and minimal working programs are provided for students so that they engage with central aspects of the computational modeling process in mechanics – working with the model parameters and initial conditions and constructing the loop which does the motion prediction. The guiding done by tutors emphasizes those aspects and encourages students to work on those aspects in their group - negotiating the code that is eventually written.

The group focus of the course coupled to the group's interactions with their tutors has helped us mitigate one of the major challenges with teaching computational modeling – keeping each student engaged with the work (i.e., not having the "VPython" person). While we have not completely solved this problem, we have gone a long way to ensuring that all students have the opportunity to work with computational modeling. The group aspect of the course is cultivated over time, that is, students learn how to work productively in their group through interactions with their tutors and the written formative feedback they receive. These two aspects emphasize to students that engagement in the work, sharing ideas, debating with their classmates, and taking responsibility over their learning are rewarded (with higher scores). The written formative feedback often identifies things that a specific student is doing well and things that this student can work on. Inevitably, one or more students in a group will receive critical feedback in the first week about their engagement with computational modeling. Sometimes that feedback is that they are driving the work too much and taking opportunities away from other students, but more often, it is that they stepped back and are letting another student control the work. The formative feedback coupled with the 3 computational projects right at the beginning of the semester emphasizes that computational modeling is a core part of the course and that we expect each student to learn it (not just the "VPython" person). By encouraging more hesitant students and pushing more engage students to act as facilitators, we have been able to mitigate the computational anxiety so common among students in these introductory courses.

Our instructional model is not without shortcomings. As the pedagogical emphasis is on the group aspect not every student performs the same tasks as every other student in the group. That is, not all students will write the computational model; there is only one computer for 4 students. But, all students will contribute to some aspect of the model - emphasizing the modeling process, which includes a fair amount of debate and discussion. And where we notice exclusion, we will address it. However, if your learning goals include that each student should write their own program, our instructional model does not do that. In addition, our scaffolds are such that students never write a computational model from scratch (i.e., starting with a blank screen). This decision was intentional as we wanted to emphasize LG 1 and LG 2 (above) in this class. We took into account where this course falls in a student's career (i.e., at the beginning) and which majors we would be instructing (i.e., mostly engineers). We decided that exposure to computational modeling with a strong emphasis on the interpretation of code and the modeling of interactions would serve the broadest population of students early in their careers. In future work, we expect that some students would have additional experiences with computational modeling where they would be held responsible for other aspects of the work. In fact, greater coordination of this course with students' future engineering course work could lead to changes to the computational emphasis in students' engineering course (i.e., being able to build off the foundation from our course).

While there are some shortcomings to our model that others might find problematic, we will emphasize that little research has been conducted on how students use computational modeling in undergraduate physics courses. The work that has been done suggests that emphasizing modeling and providing significant scaffolding at the introductory level is important for student success [9,23]. However, the details of how this work is done and what impact it has on student learning, motivation, and affect is understudied. In part, the $P^3$ course is a context for research into these and other areas. As we uncover new research results, we will share those with both the physics education research community and the physics education community writ large as we all begin to navigate teaching students to use the tools of 21st century science.